\documentclass[prb,aps]{revtex4}
\usepackage{amsmath}
\usepackage{epsf}
\usepackage{color}
\usepackage[dvips]{graphicx}
\usepackage{amssymb}
\usepackage{pstricks}
\begin{document}



\title{Time resolved scattering relaxation mechanisms of microcavity polaritons}

\author{F. Chaves and F. J. Rodr\'{\i}guez\footnote{Email address: frodrigu@uniandes.edu.co}}
\affiliation{Departamento de F\'{\i}sica, Universidad de Los
Andes, A.A. 4976, Bogot\'a D.C., Colombia.}
\begin{abstract}
We study the polariton relaxation dynamics for different scattering mechanisms
as: Phonon and electron  scattering procesess. The relaxation
polariton is obtained at very short times by solving the
Boltzman equation. Instead of the well-known relaxation process by phonons,
we show that the bottleneck effect relaxes to the ground
state more efficiently at low pump power intensity when the
electron relaxation process is included. In this
way,  we clearly demonstrate that different 
relaxation times exist, for which any of these two mechanism
is more efficient to relax the
polariton population to the ground state.
\end{abstract}
\pacs{ 73.21.Hb\sep71.10.Pm\sep73.23.Ad\sep73.50.Jt}
\maketitle
\section{Introduction}
Many interesting and unusual dynamical
quantum many-body phenomena have been the subject of an extended
study. In particular, using ultrafast laser technology, the
 dynamics of many-body processes can be understood from a fundamental
point of view. Extended reports in the literature suggest that due
to the strong coupling between  radiation-matter, signatures of a
possible condensation of bosonics particles can be detected. One
of the most spectacular and still controversial effects is the
possibility to get Bose Eintein Condensation of polaritons in
quantum wells embedded in semiconductor microcavities. The
accumulation in the ground state of polaritons has been suggested
by different
experiments\cite{savvidis,lagodakis1,stevenson,baumberg} and
predicted theoretically by\cite{ciuti,rubo,cao}. By illuminating
the sample with lasers of different frequencies, it is possible
to get a matter-wave amplification without requiring a coherent
pump-laser\cite{senellart,huang} However, one of the main
difficulties to obtain a Bose condensation of polaritons has been
the relaxation of the bottleneck which hinders that excitons relax
quickly to the optically active lower energy states. This
bottleneck effect, can be relaxed if the pump-laser is increased
 or if different relaxation scattering mechanisms become more
efficient to allow that the population relaxes to the ground
state\cite{lagodakis,malpuech}.

A comprehensive study of the main mechanisms of polaritonic
relaxation, comprises electron, phonon and polariton-polariton
scattering. In the stationary limit, Malpuech et al\cite{malpuech},
 have shown results for each of the above mentioned 
scattering mechanism for different pump laser intensities. As can
be experimentally observed, for some range of carrier densities,
where the strong-coupling remains, the polariton-polariton
relaxation is very inneficient\cite{butte,sermage} to overcome the
bottleneck. However, in semiconductor microcavities, under
non-resonant excitation, acoustic phonons provide an
effective channel for polaritons relaxation\cite{savidis1,thoai}.
Moreover, recent experiments\cite{savvidis,tartavoski} have shown
the importance of electron-polariton scattering at densities where
the electron-electron interaction are insufficiently effective due
to the their fast relaxation. This two relaxation
mechanisms could be the most relevant to relax the huge population
at the bottleneck. In spite of the wide amount of experimental and
theoretical\cite{carlos,haug} studies on this subject that has
appeared, theoretical and detailed calculations as a function of
time are very few and consider only one type of relaxation
processes. The case in which both electron- and phonon-polariton
interactions are present has only been studied recently in the
stationary limit\cite{savvidis} either for high laser power
excitations or weak detuning regimes. In our opinion one of the
important questions in the area of polariton relaxation dynamics
is the effect of the different relaxation mechanisms in the
transient time range. It has been recently suggested that in the
transient time and in an isotropic\cite{thoai} system, an optimal
relaxation bottleneck can be achieved only by increasing the power
laser excitation. In this paper, we show that this result becomes
ambiguous and, as is well known, in the stationary limit, is
partially true in a restricted range of power excitation.

We report in this work on the dynamics of polaritons continuously excited
by a laser and including both electron- and acoustic-phonon scattering process.
This introduces
quantitative and qualitative different features for the dynamics following
the resonant excitation. 
 The purpose of the present paper
is to shed light about which is the more efficient relaxation
mechanism for polaritons for a given pump laser power and for a finite
electron concentration, as a function of time and momentum.
 The main result is that electron-polariton
(el-pol) relaxation is the most efficient mechanism, when 
compared with the phonon-polariton (ph-pol) relaxation. The road
to studying the contribution of different scattering mechanisms
has lead to use one of the more accepted models based on solving
the Boltzmann equation, because in this way it is very easy to
include different scattering mechanisms and capture the essential
physics involved in the polaritonic relaxation process.

The paper is structured as follows: Section \ref{hi} describes
the Hamiltonian interaction between excitons and photons. Sec \ref{relax} 
describes the rate equation and the scattering process
including both phonon and electron scattering. Section
\ref{numerical} shows explicitly how our numerical simulation
gives the temporal range in which either scattering mechanisms are
more important. Finally we present discussions and conclusions of
our results and establish the comparison with another theoretical
result.

\section{Theoretical Model}
\label{theomodel} We describe first in this section the model that
we use in order to describe the polariton relaxation. We assume through 
this work that the polaritons
interact with phonons through a deformed and not a piezoelectric
potential.

\subsection {Interacting Hamiltonian}
\label{hi} The usual experimental set-ups consider a quantum well
embedded in a microcavity which allows a stronger coupling between
the two dimensional excitons and the confined light. The excitons
are generated by pumping at lower excitation and an attractive
Coulomb interaction arises, showing an spectrum similar to the
hydrogen atom. The spectrum of the bound exciton states in a
quasi-two dimensional system is given by
\begin{eqnarray}
\omega_e(\vec k)=-\omega_e(0)+\frac{k^2}{2M}
\end{eqnarray}
where $M=m_e+m_h$, $\omega_e(0)=\frac{m_x e^4}{2\epsilon_s^2}$, $m_x=m_e m_h/M_x$, 
 $\epsilon_s$ is the effective
dielectric semiconductor constant and $m_{e(h)}$ represents the effective
electron (hole) mass of the carriers in the semiconductor, $(\vec k)$
the in-plane momentum vector ($\hbar=1$).
In this way the total Hamiltonian can be written as:
\begin{eqnarray}
H=\omega_{cav}(\vec k)c_{\vec k}^\dag c_{\vec k} +\omega_{e}(\vec k)
e_{\vec k}^\dag e_{\vec k}+
\Omega_{0}(c_{\vec k}^\dag e_{\vec k}+e_{\vec k}^\dag c_{\vec k})
\end{eqnarray}
where $\omega_{cav}(\vec k)=\frac{c}{n}{\sqrt{k^2+\frac{(n\omega_c)^2}{c^2}}}$,
 is the resonant cavity energy,
$\omega_{e}(\vec k)=\omega_{e}(0)+\frac{ k^2}{2M}$
is the exciton dispersion enrgy, $\omega_c=\frac{\pi c}{n L}$, $n$ is the refraction
index of the semiconductor and $L$ the effective well width.
The photon(exciton) creation operators are given by
$c_{\vec k}^\dag(e_{\vec k}^\dag)$ and $\Omega_{0}$
is the coupling energy between the excitons and photons.
By using a linear transformation the exciton-photon interaction
Hamiltonian can be diagonalized which
leads to new quasiparticles, called excitonic-polaritons\cite{tasone}. This
new quasiparticles arise by the mixing of the exciton and photon
states. It is well known that at very low excitonic
densities, the excitons follow Bose commutation relations,
given that the excitons can be considered as the excited state of
a N-particle system, in such a way that the excitons can be
considered as bosonic particles. The Hamiltonian
can be diagonalized to obtain the well known branches of the
polaritons dispersion relation\cite{tasone}
\begin{eqnarray}
E_{L}^{U}(k)=\frac{\omega_{cav}(\vec k)+ \omega_{e}(\vec k)}{2}\pm
{\sqrt {\Omega_{0}^2+\frac{\Delta^2(\vec k)}{4}}}
\end{eqnarray}
$\Delta(\vec k)=\omega_{cav}(\vec k)-\omega_{e}(\vec k)$ represents the
detuning between the exciton and cavity energies. For negative detunings ($k=0$)
a resonance condition can be achieved for higher in plane wave
vectors.

\subsection {Phonon and electron relaxation mechanisms}
\label{relax} In order to describe better the experimentalist
results\cite{lagodakis}, we describe in detail the two most
important relaxation mechanisms of polaritons to the lowest energy
states for lower power excitations: {\sl (1) Acoustic phonon
relaxation} and {\sl (ii) Electron relaxation}. Using the
Boltzmann equation, the polariton relaxation rate is given by the
following master equation:
\begin{eqnarray}
\frac{dp_k^i}{dt}=G_{\vec k}^i(t)-\frac{p_k^i}{\tau_{\vec k}^i}+
(p_{\vec k}^i+1)\sum_{j,\vec k'}W_{\vec k',\vec k}^{j,i} p_{\vec k'}^j
-p_{\vec k}^i\sum_{j,\vec k'}W_{\vec k,\vec k'}^{i,j}(p_{\vec k'}^j+1)
\label{boltzman}
\end{eqnarray}
where $p_{\vec k}^i$ is the polariton distribution of the $i^{th}$
branch, $G_k^i(t)=G_0 Th (\frac{t}{T})$ (in the following $T=20
ps$) is the continuous optical pumping generation, $\tau_{\vec k}^i$
is the time recombination rate. Due to the radiative and non
radiative character of the polaritons, the radiative lifetime is a
function of $k$. We have used the same values as in Ref.
\cite{thoai}.
  The total scattering rate (phonons + electrons) is denoted
by $W_{\vec k,\vec k'}^{i,j}=W_{\vec k,\vec k'}^{i,j (ph)}+W_{\vec
k,\vec k'}^{i,j (e)}$ and describes the transitions between two 
polariton states.
It is worth pointing out that the algebraic properties of the
above equation guarantee that its solution will consist of a
linear combination of terms with a decaying exponential
time-dependence, and will always show a stable approach to some
steady state and it is appropriate for the description of the dynamics for
low-density excitation.

-{\sl Acoustic phonon-polariton relaxation:} For resonantly or non-resonant
created excitons , the transition rates between two polaritonic
states are proportional to the matrix elements between exciton
states weighted by the respective exciton content of the polariton
states.
\begin{eqnarray}
W_{\vec k,\vec k'}^{i,j, (ph)}=2\pi\sum_{q_z}
|M^{i,j}_{ph}(\vec k-\vec k')|^2 (n(\Omega_{\vec q,q_z})\pm \frac{1}{2}+\frac{1}{2})
\\
\nonumber
\delta(E_j(\vec k')-E_i(\vec k)\pm \Omega_{\vec q,q_z})
\end{eqnarray}

where $M^{i,j}_{ph}(\vec k-\vec k')$ is the matrix element of the
deformation potential between polaritons states in different
branches $i,j$, given in Ref. \cite{tasone}. $E_j(\vec k')$ is
the polariton energy in the branch $j$ with wave vector $\vec k'$
in the two dimensional quantum well plane, $\Omega_{\vec q}$ is
the acoustic phonon energy with $\vec Q= (\vec q,\vec q_z)$ the
total phonon momentum and $n(\Omega_{\vec q})$ is the Bose
occupation number for phonons.

-{\sl Electron-polariton relaxation:} 
When polaritons interact with electrons, the experimental and
theoretical results show that this mechanism becomes more
efficient in the stationary regime. The advantages of including
the present relaxation process relies on the fact that low
electron density can be tuned easily. It is important to 
remark that the advantage of this scattering mechanism 
with respect to the previouly discussed one, is that as 
the electron has a very small effective mass than an 
exciton, and the relaxation of bottleneck 
polaritons to the ground state requires few 
 scattering processes.
The matrix elements for the transition rates are calculated as:
\begin{eqnarray}
W_{\vec k,\vec k'}^{i,j (e)}=2\pi\sum_{\vec k_e}
|M^{i,j}_{(el)}(\vec k-\vec k')|^2 f_{\vec k_e} (1-f_{\vec k_e+\vec k-\vec k'})\\
\nonumber
\delta(E_j(\vec k')-E_i(\vec k)-\frac{1}{2 m_e} (k_e^2-|\vec k_e + \vec k -\vec k'|^2))
\end{eqnarray}

where $f_{\mathbf{k}_e}$ is the Fermi-Dirac distribution function for an
${k}_e$ electron momentum
and $M^{i,j}_{(el)}=V_{dir}\pm V_{exc}$ is the matrix element of interaction
between an electron and polariton\cite{savvidis}, corresponding to the triplet
configuration (+, parallel electron spins, $V_{dir}$), and the single
configuration (-, antiparallel electron spins, $V_{exc}$). In the following, we shall
only include the stronger electron-polariton scattering of triplet
configuration.The electron Fermi-Dirac distribution is given by
$ f_{k_e}=[e^{(E_e(k_e)-\mu)/k_B T}+1]^{-1}$
where in the two dimensional case the chemical potential  is
$\mu=k_BT \ln(e^{E_f/K_BT}-1)$, and we assume that the Fermi
energy is given by the free 2D electrons value:
$E_f=\pi n_e/m_e$,  $n_e$ being the free electrons density. We also
assume that the dynamics of free electrons is so fast that we may 
assume that equilibrium is reached at times shorter than other ones.\\

\section{Numerical results}
\label{numerical} In order to clarify how is the effectiveness of
ph-pol or el-pol,
 we compare the results for both mechanisms on the
relaxation polaritons.  The Bolztmann equation [\ref{boltzman}]
includes both electron and phonon relaxation. The details of the
$W$ matrix elements have been described in detail in
Ref.\cite{savvidis}. We present results for the following
set of parameters:
 Well width of $L_z=60\dot A$, Rabi splitting $\Omega=4.8 meV$, $n=3.07$,
electron (hole) deformation potential $D_{e(h)}=12 (-7) eV$,
longitudinal sound velocity $u=4.81 \times 10^3 m/s$, mass density
of the solid $\rho=5.3\times10^3 kg/m^3$, radiative
cavity(exciton) times $\tau_{c(x)}=50 (200 ps)$
The decay and generation rates as a function of $k$ are
taken from Ref.\cite{thoai}.

In Fig. 1(a), 1(b) and 1(c) we show the two possibilities for relaxation
mechanisms. Fig. 1(a) corresponds to the polariton population
relaxing by acoustic phonons and in Fig. 1(b) the same but only
including el-pol relaxation. Fig. 1(c) includes at the same time
both mechanism. We have taken $T=2 K$,
$\delta=\omega_{cav}-\omega_{exc}=-1 meV$. As can be seen, the
bottleneck suppression occurs due to the el-pol effects, but it is
not lifted completely. In Fig. 1(a) the polariton population reaches its
maximum through phonon relaxation for $k\approx 1.5 \times 10^{4}
cm^{-1}$ while for el-pol relaxation it start to rise at $k\approx
3.2 \times 10^{4} cm^{-1}$ as can be seen in Fig. 1(b). The
differenice in the position of the maximums is a consequence of the restricted
phase space filling at this low temperatures, where emission
polariton processes are less probable than absorption processes.
However, for $t<0.5 ns$, a polariton population grows due to the
relaxation by phonons and the relaxation by free carrier electrons
starts to grow after 0.7 ns. In Fig. 1(c) both mechanisms are
included at the same time. The present results show that for this low
temperatures it is impossible to distinguish which of both
relaxation mechanisms is relevant to suppress the bottleneck. It
is important to remark that the bottleneck is relaxed if the
non-resonant power pump is increased
 and the electron relaxation
becomes more inefficient than phonons (not shown).

In order to describe better the experimental data, and to
elucidate which of these mechanisms is more important, the lattice
temperature is raised to $T=10 K$ with the same parameters as
in Fig. 1. The numerical data are depicted in Fig. 2(a)
(ph-pol), Fig. 2(b) (el-pol), Fig. 2(c) (ph-pol + el-pol). As
 be expected, the ph-pol scattering allows the bottleneck
relaxation to lower $k$ states, but the bottleneck is partially
depleted and remains at $k\approx 1.2 \times 10^{4} cm^{-1}$,
showing a complete range of empty states below this value.
However, if the el-pol mechanism is included, the bottleneck relaxes to
lower, states showing that it becomes more efficient than ph-pol.
 In Fig. 2(b) it is possible to see that the el-pol mechanisn
becomes more efficient to relax completely the bottleneck and
populates the $k=0$ states. When both mechanisms are included,
clearly the restriction of the state space is broken and the
lower states becomes polariton populated. Our results explicitly
show that for lattice temperatures greater than the electronic
temperature ($k_BT_e=\pi n_e/m_e= 2 K$) the electron gas becomes
the more efficient mechanism to the polariton relaxation. The
population reaches the steady state around $t=1.5 ns$, and after this
time the phonon relaxation becomes more inefficient.

In Figs. 3. we show the polariton population distribution for
different times. Fig. 3(a) shows that for $T=2 K$ the
populations follow a Boltzmann distribution and for both
relaxation mechanisms the steady state is obtained at $t\approx 2.5
ns$. Fig. 3(b) includes both mechanism showing that el-pol
relaxation is not efficient enough. Moreover, if the temperature
is increased, $T=10 K$, in Fig. 3(c)  the polariton population maximum peak
is exchanged with respect to Fig. 3(a) and lower $k$ states
become populated,
 showing clearly an asymmetric behavior for
the polariton population with respect to the $k-$ value where the
population takes its maximum. Clearly, for low excitation, the shape of the
population develops a symmetric shape for low temperatures, but it becomes deformed when
the temperature is increased, demonstrating clearly that the main important
relaxation mechanism for the bottleneck comes from electron
scattering.
 Our results are in better agreement with the experimental
results shown in Refs.\cite{butte,tartavoski}.
Allowing the el-pol scattering, the bottleneck is shifted towards
lower momenta, changing the population in an asymmetric way
compared with the pnonon-polariton relaxation.

In summary, we have used a model of bosonic particles with a
strong electron-polariton and phonon-polariton coupling, and have
shown numerically exact results for the population relaxation as a
function of momentum and time. The competition between the
itinerancy of the electron and the phonon interaction gives rise
to a changeover of the behavior of the polariton population from
low temperatures to high temperatures in which we elucidate which
one of these two mechanisms is more important in order to relax
the bottleneck. Our results demonstrate clearly that Doan's\cite{thoai}
theoretical results are valid for very low temperatures and
explain scarcely the experimental data. Our results, additionally show that
there exists a fundamental difference between a relaxation involving 
electron or acoustical phonon elastic scattering with polaritons.
We have clearly shown that the electron-polariton mechanism
becomes more efficient for $t\approx 2 ns$ and it is
the responsible for the relaxation in the stationary limit, leading to a huge
population at $k=0$ values.

The authors acknowledge to L. Quiroga and A. Reyes for fruitful discussions. Financial
support from Facultad de Ciencias, Universidad de los Andes, Banco de la
Rep\'ublica and COLCIENCIAS:1204-05-11408 are gratefully acknowledged.

\begin{figure}[h!]\label{xe1}
  \includegraphics[scale=0.5]{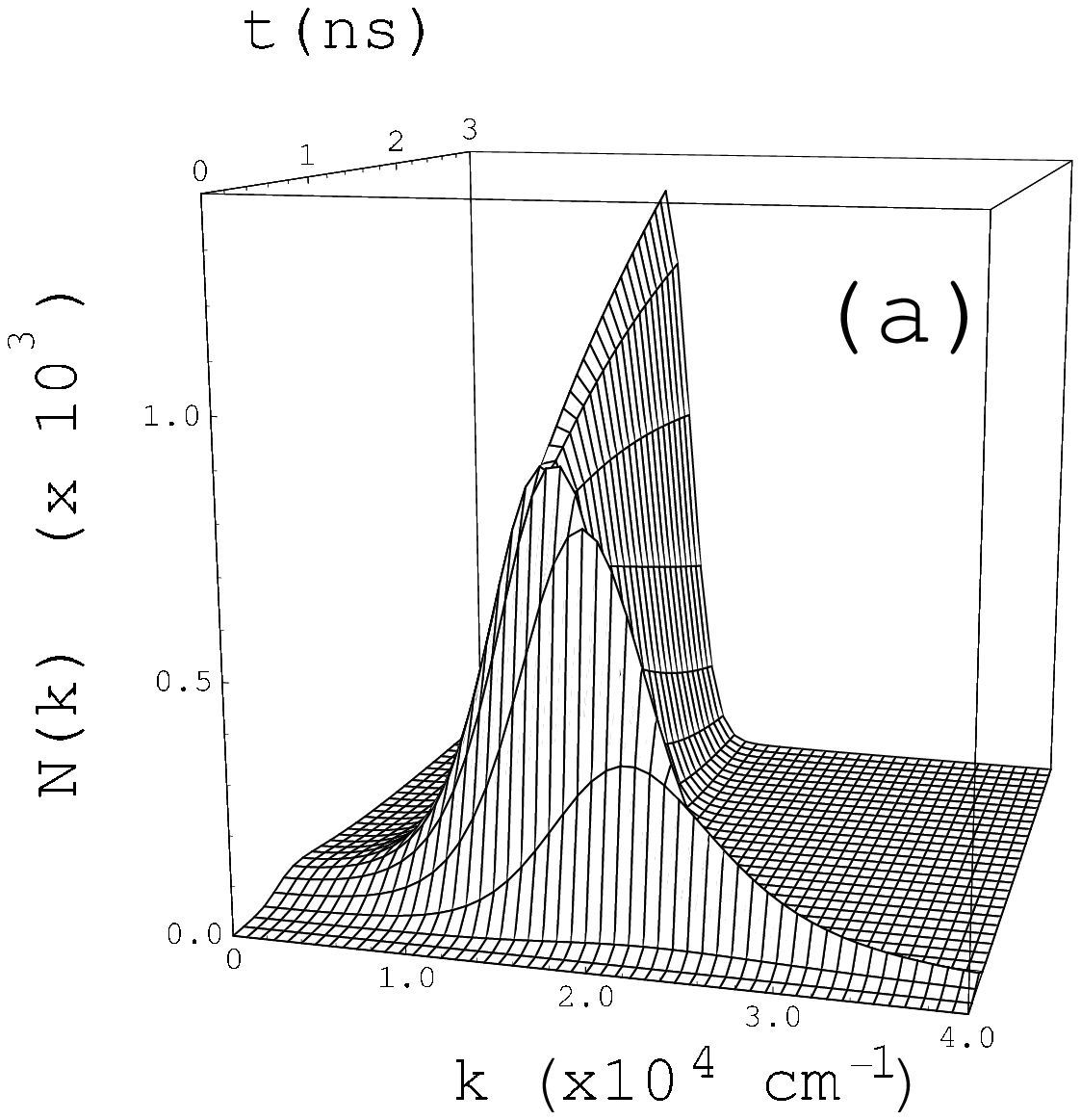}
  \includegraphics[scale=0.5]{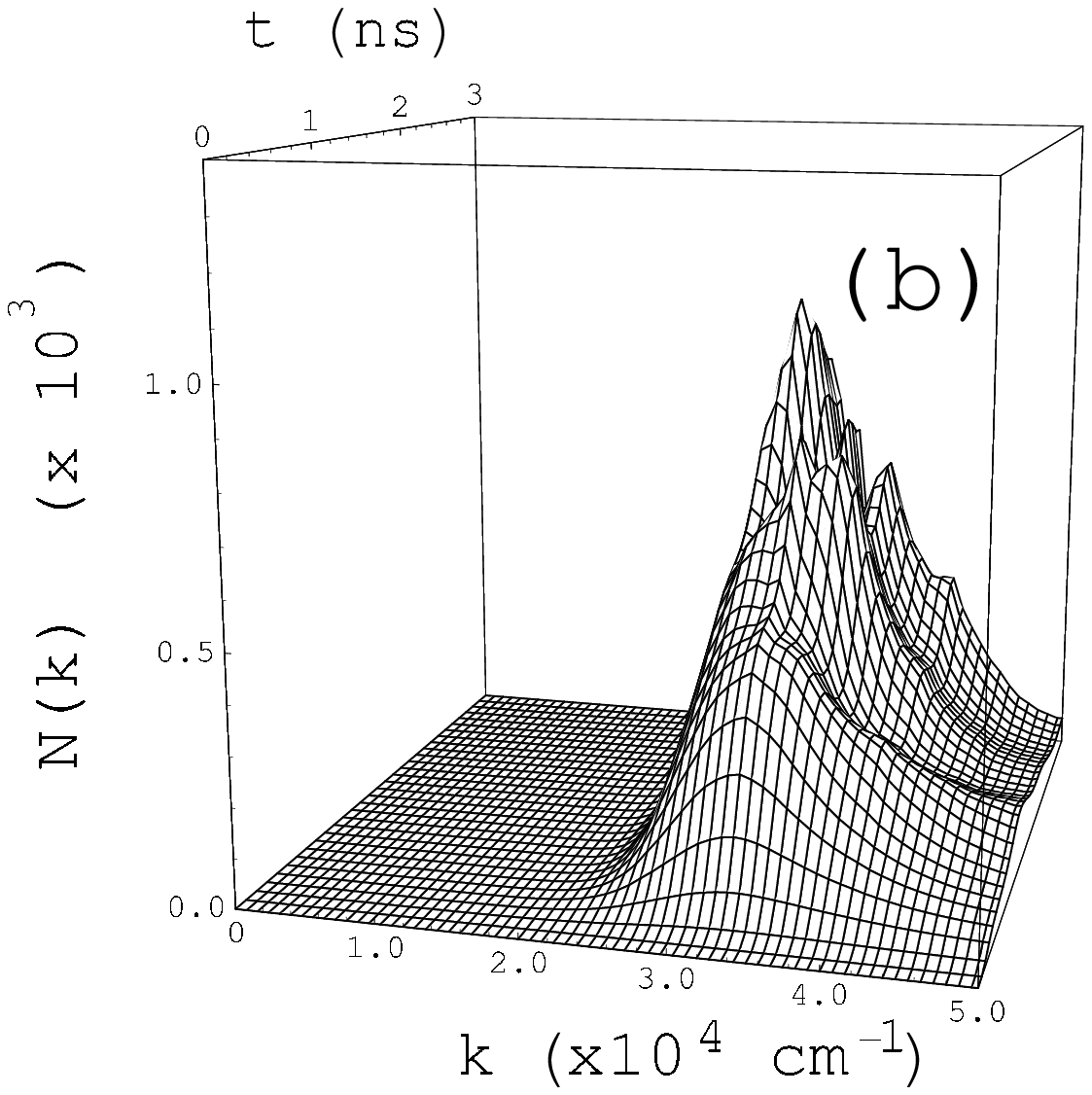}\\
  \includegraphics[scale=0.5]{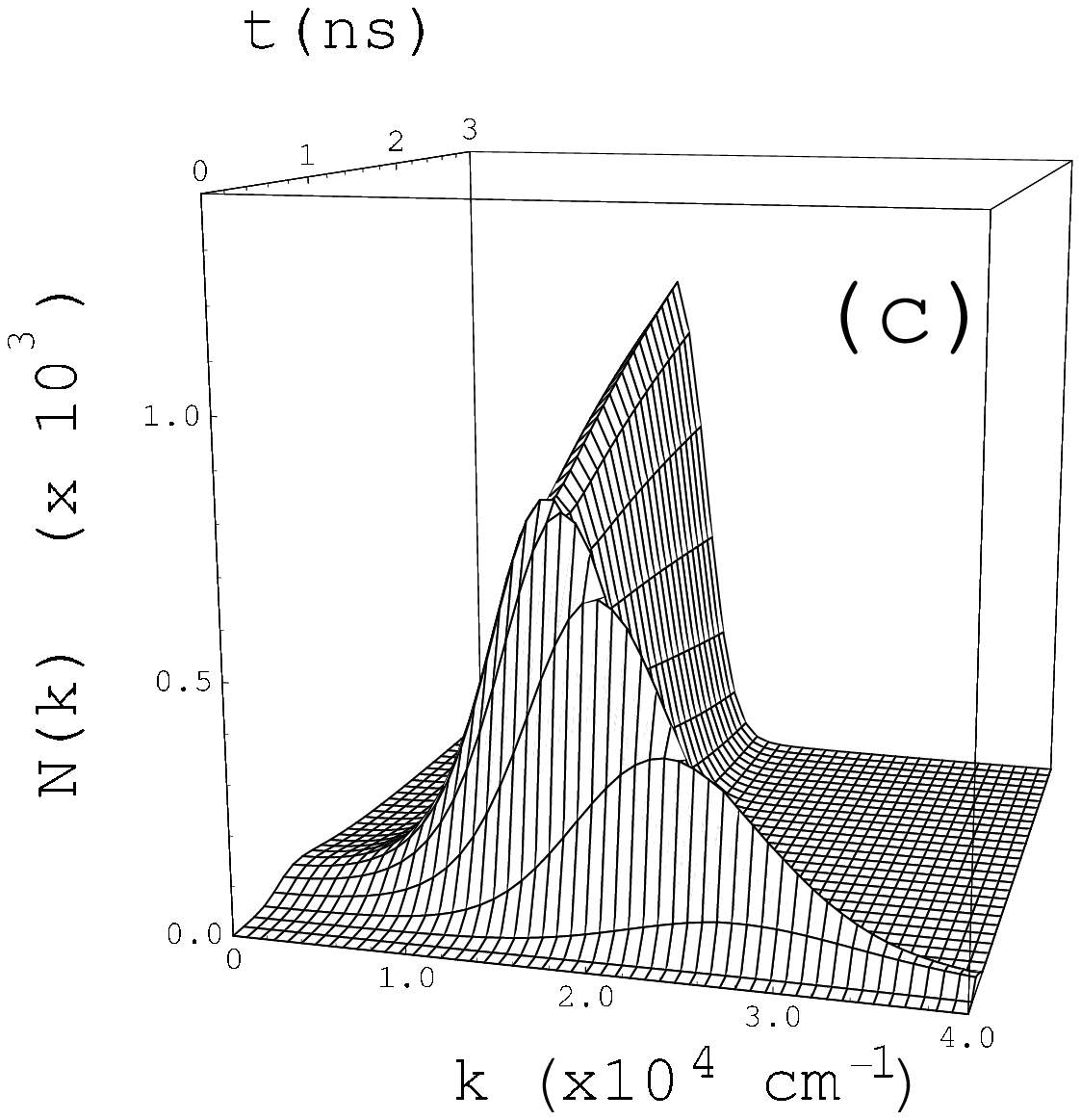}
  \caption{Distribution function for different interaction mechanisms with
   $n_x=8\times 10^9\,cm^{-2}$. $T=2\,K$, $\delta=-1\,meV$ and the
pump laser is non-resonant applied at $k=6\times 10^4 cm^{-1}$. a) Only
ph-pol relaxation; b) Only el-pol relaxation;
  c) Both mechanism includes at the same time}
\end{figure}

\begin{figure} [htbp]
\centering
\includegraphics[height=8.5cm,width=6.8cm]{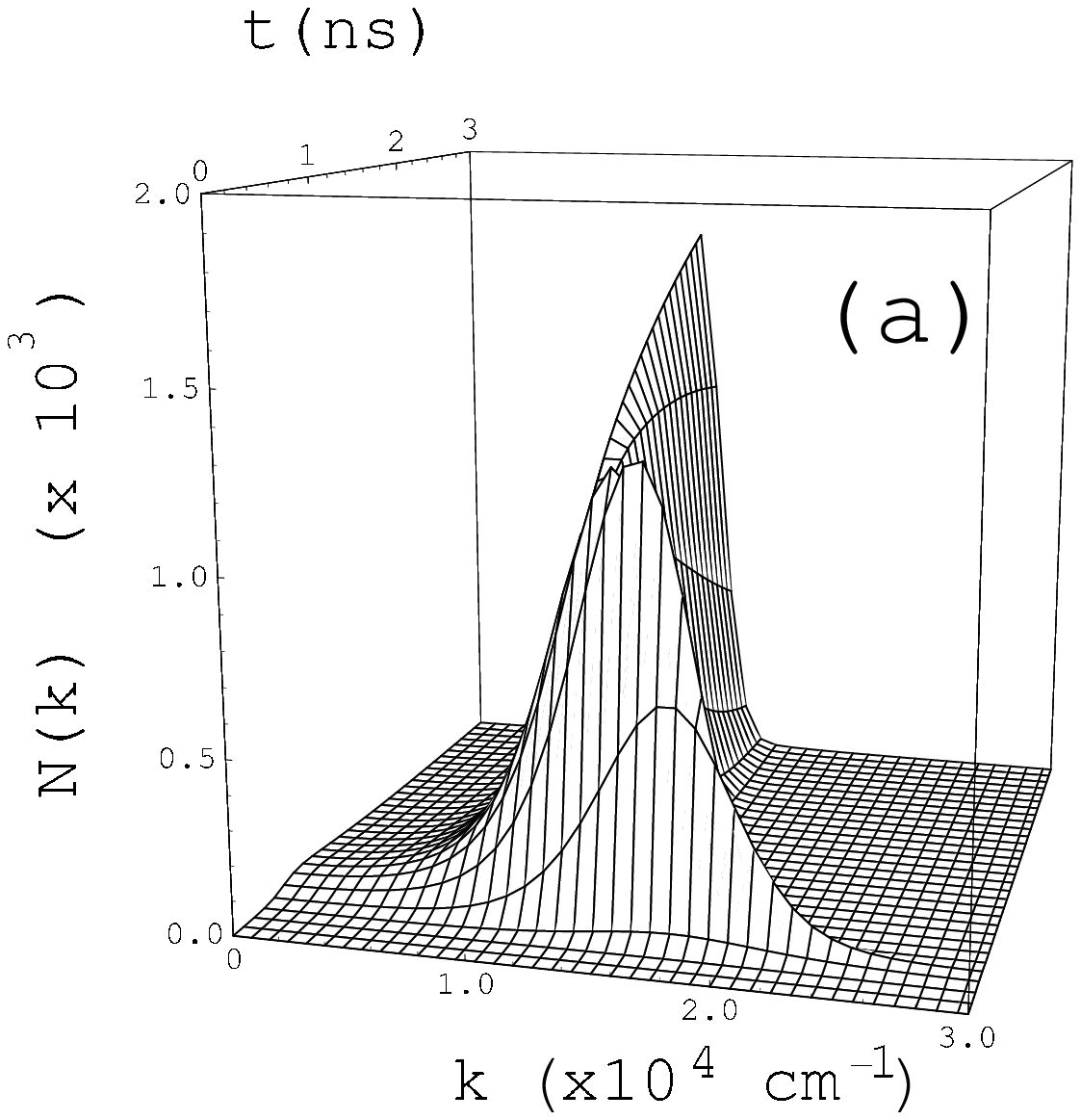}
\includegraphics[height=8.5cm,width=6.8cm]{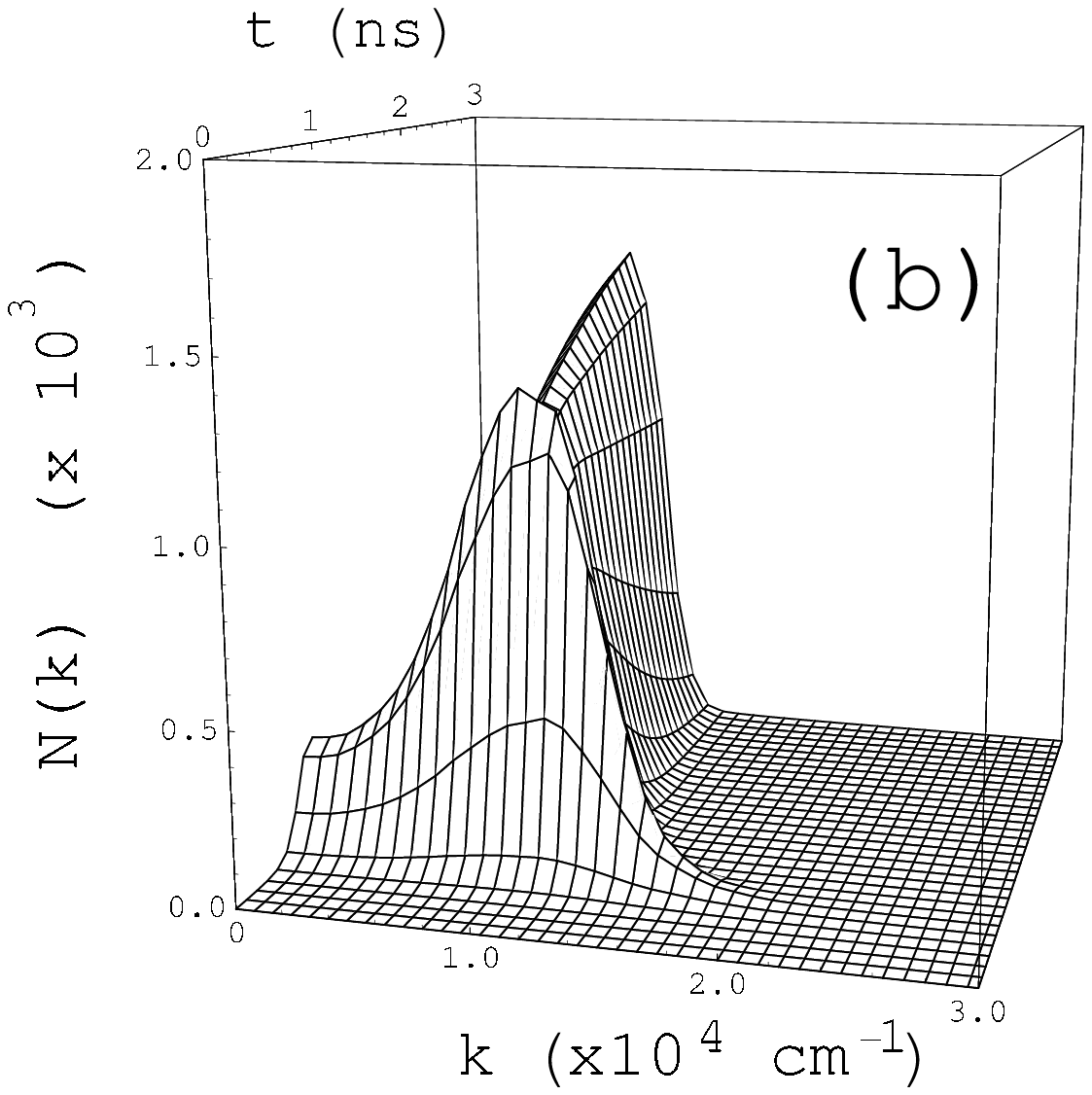}
\includegraphics[height=8.5cm,width=6.8cm]{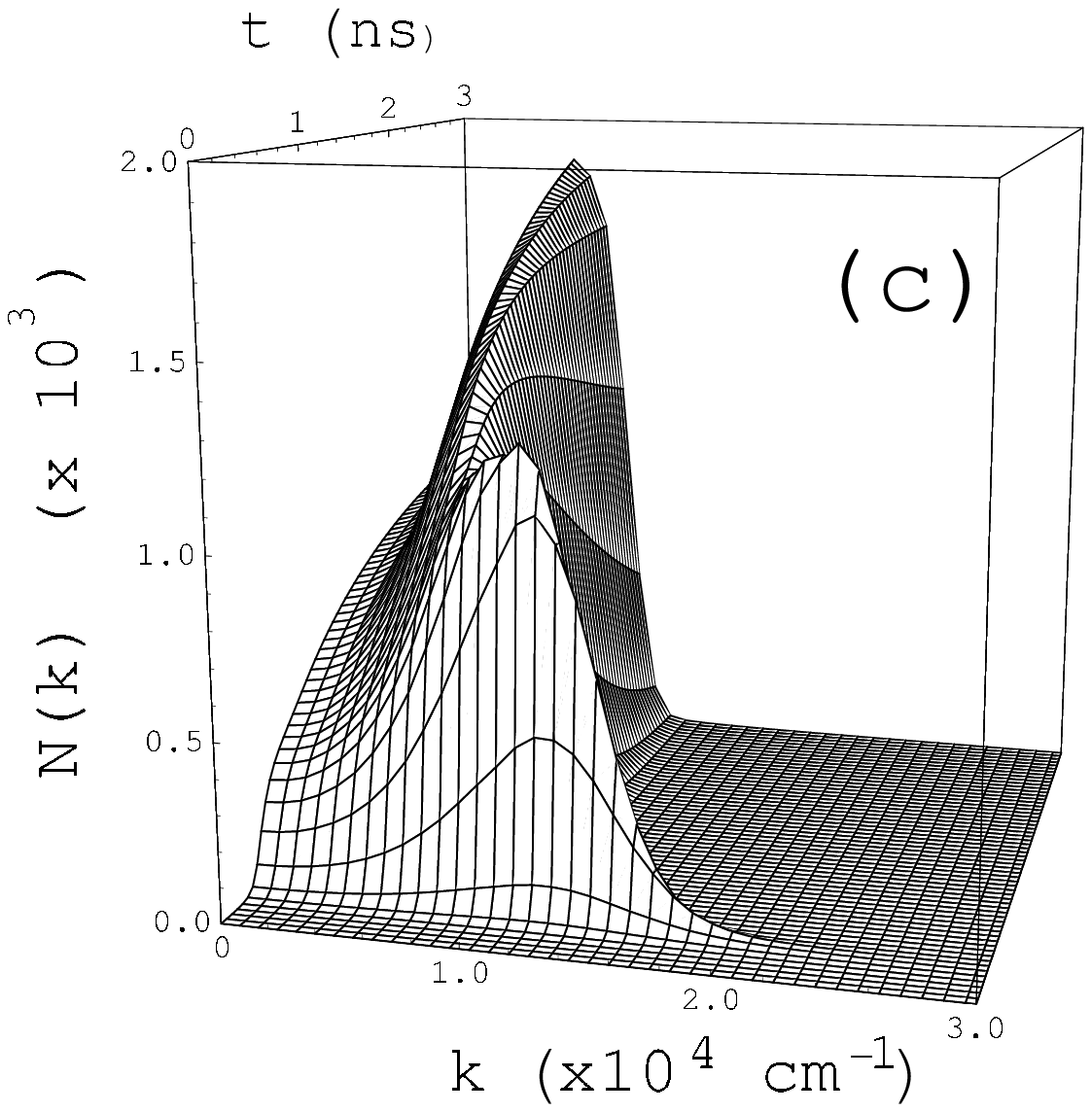}
  \caption{Distribution function $N_k(t)$ with $T=10\,K$, $n_x=8\times 10^9\,cm^{-2}$ and
  $\delta=-1\,meV$ for
  a) Only ph-pol interaction mechanism,
  b) Only el-pol interaction mechanism,
  c) ph-pol + el-pol interaction mechanisms.
  }
  \label{fig:fig1}
\end{figure}

\begin{figure}[h!]\label{totalT=10}
 \centering
  \includegraphics[angle=-90,width=6cm]{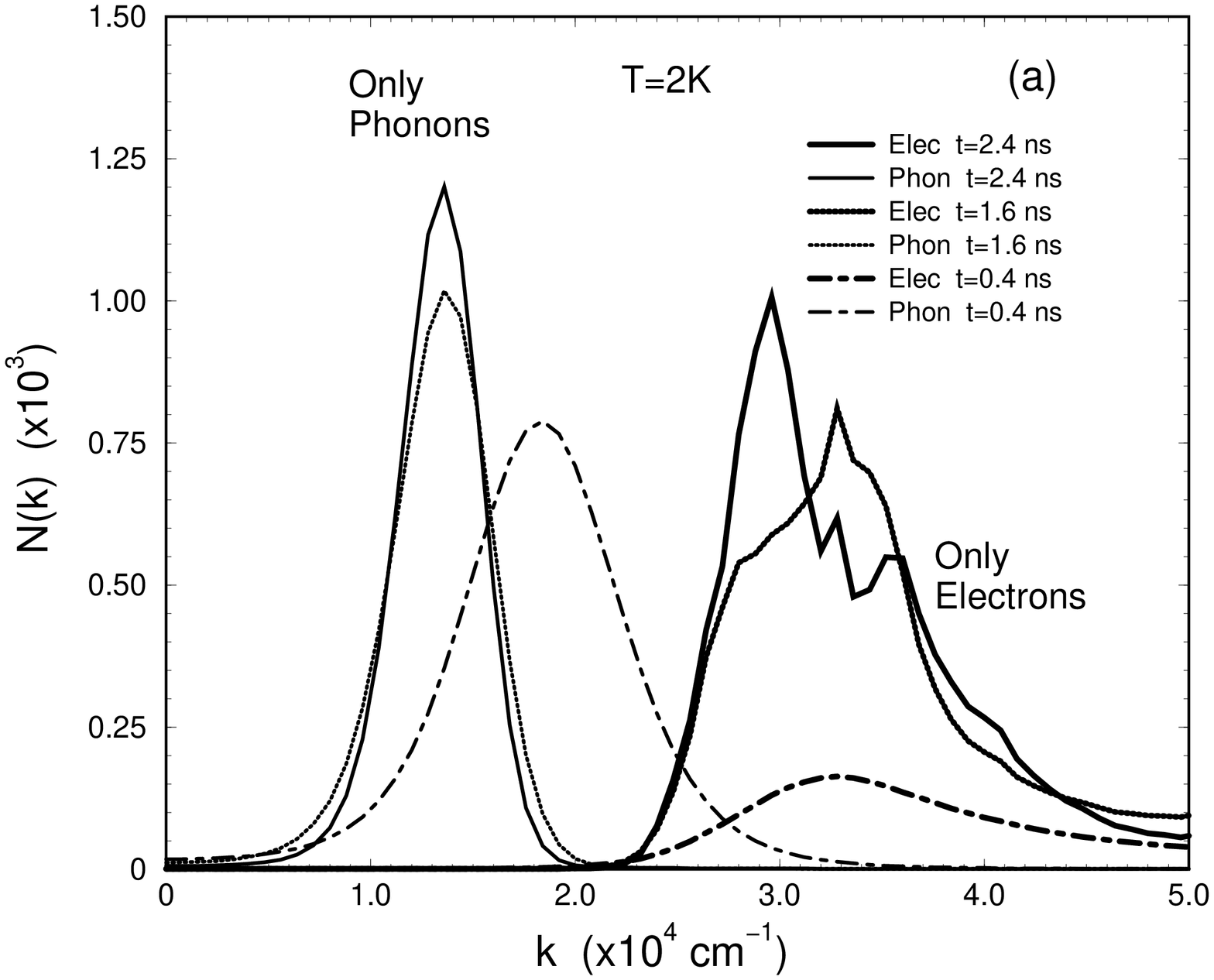}
  \includegraphics[angle=-90,width=6cm]{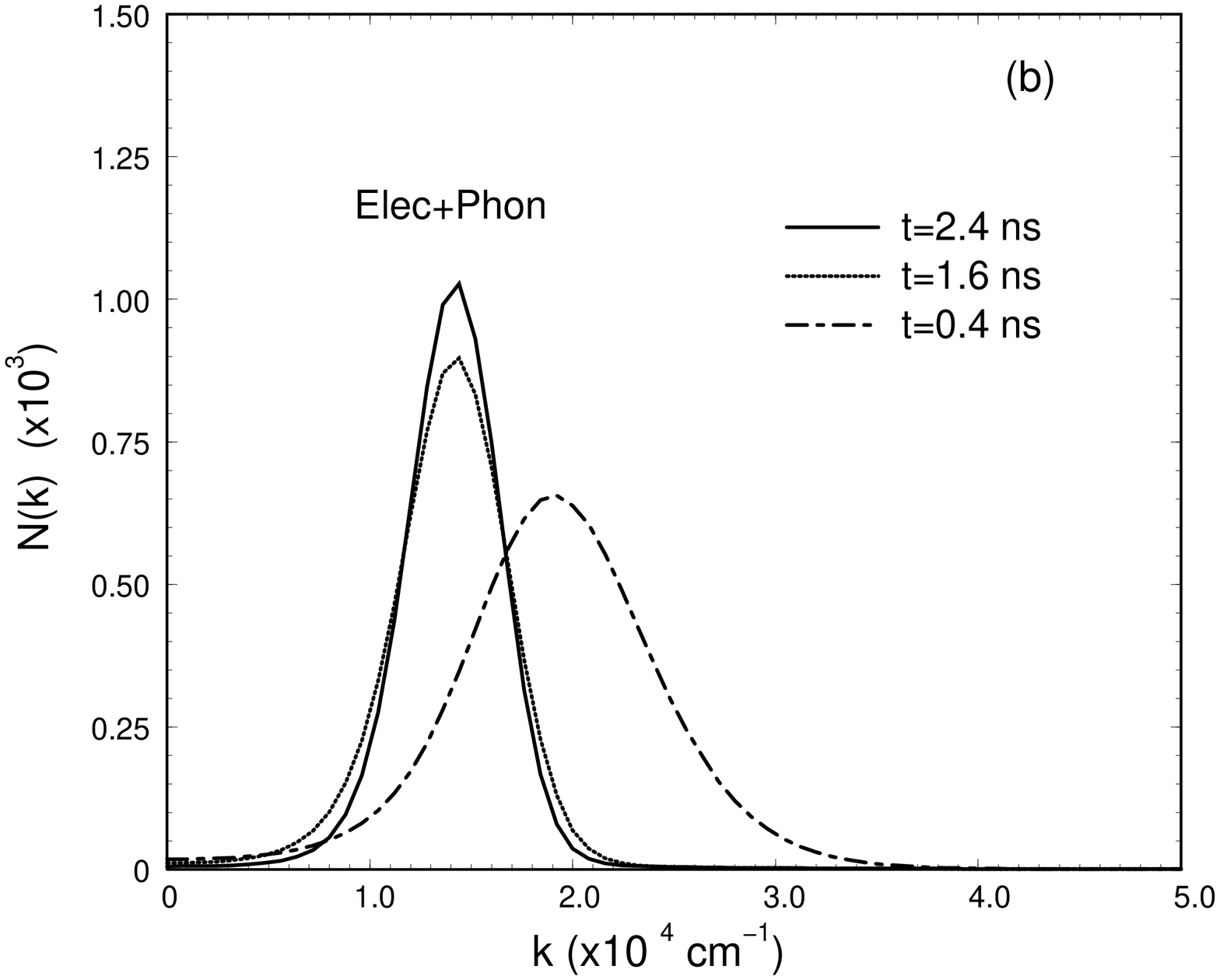}
  \includegraphics[angle=-90,width=6cm]{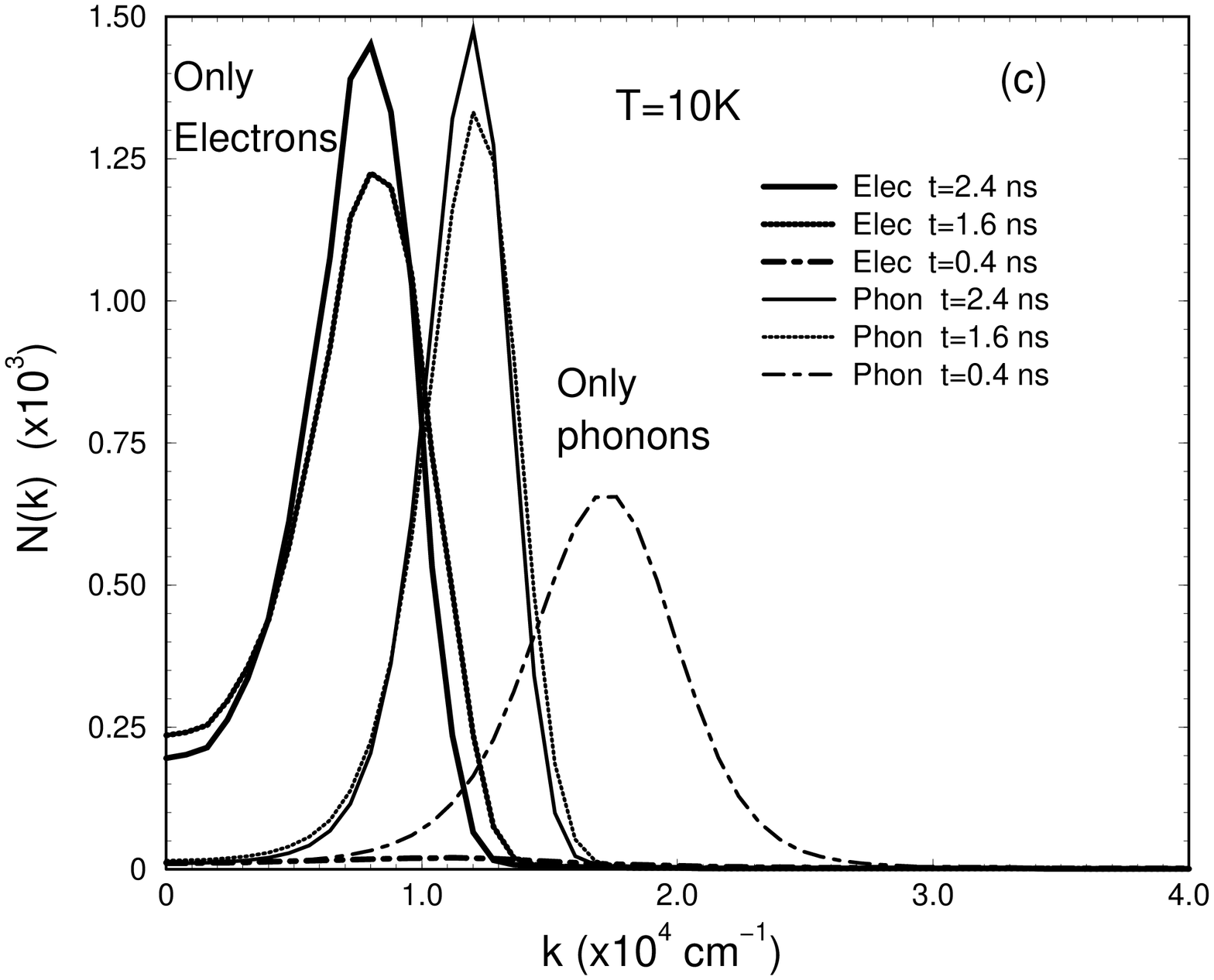}
  \includegraphics[angle=-90,width=6cm]{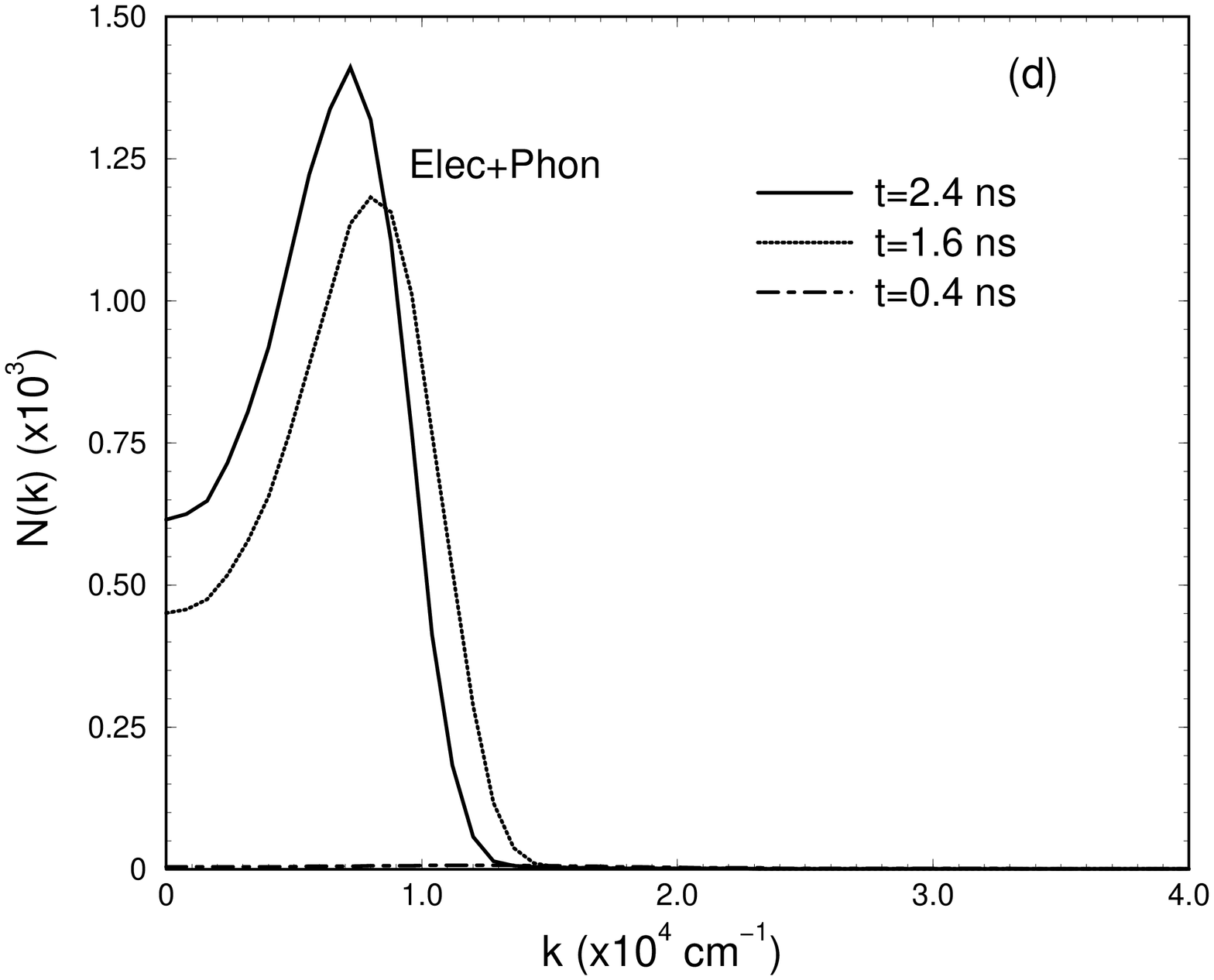}
  \caption{Comparision of the time evolution of the polariton population at T= 2K relaxating
through (a) phonons and electrons, alone each one  and (b) the sum of
both processes. (c) and (d) the same as (a) and (b) but for T=10 K up to 2.4 ns}
\end{figure}

\end{document}